\documentclass{article}
\usepackage{cite}
\usepackage{graphicx}
\usepackage{dcolumn}
\usepackage{amsmath}

\input{tcilatex}
\begin{document}

\date{}
\title{On the quantum-mechanical singular harmonic oscillator}
\author{Francisco M. Fern\'{a}ndez\thanks{%
fernande@quimica.unlp.edu.ar} \\
INIFTA, DQT, Sucursal 4, C.C 16, \\
1900 La Plata, Argentina}
\maketitle

\begin{abstract}
We obtain the eigenvalues and eigenfunctions of the singular harmonic
oscillator $V(x)=\alpha/(2x^2)+x^2/2$ by means of the simple and
straightforward Frobenius (power-series) method. From the behaviour of the
eigenfunctions at origin we derive two branches for the eigenvalues for
negative values of $\alpha$. We discuss the well known fact that there are
square-integrable solutions only for some allowed discrete values of the
energy.
\end{abstract}

\section{Introduction}

\label{sec:intro}

Exactly solvable models are most useful for teaching some of the subtleties
of quantum mechanics in introductory courses. For this reason most textbooks
on quantum mechanics\cite{CDL77} and quantum chemistry\cite{P68} discuss at
least the harmonic oscillator and the hydrogen atom. The former is useful
for the analysis of the vibrational spectra of diatomic molecules and the
latter for an introduction to atomic physics. The Schr\"{o}dinger equation
for these simple models can be solved in many different ways, one of them
being the power-series method (also called Frobenius method\cite{BO78}).
From time to time teachers propose other quantum-mechanical models that
exhibit particular features that may not appear in the problems just
mentioned. For example, the so called isotonic oscillator, pseudoharmonic
oscillator, singular harmonic oscillator or modified harmonic oscillator has
received considerable attention\cite
{D32,GK61,WJ79,S84,SG85,B88a,B88b,SE88,S89,B89,BS90,BD90,KGS90,
BDO92,L94,P95,P98a,P98b,P99,P01,WGMD02,DM02,PR03,S03,SHV03,DSL05,PRML05,R05,SHK05, SD06,STAY07,PS07,DMG07,IS07a,IS07b,IS08,OAA08,C08,TS09,AS12,OS12,PD13,ND16,BB22,BBSM23}
as a model for the study of the vibration-rotation spectra of diatomic
molecules\cite{D32,S84,SG85,IS07b,AS12,OS12} or for a pseudodot\cite{C08}.
The Schr\"{o}dinger equations with this potential in any number of
dimensions can be solved exactly in terms of generalized associated Laguerre
polynomials\cite
{D32,S84,SG85,B88b,WGMD02,DSL05,OAA08,C08,AS12,OS12,PD13,ND16,BBSM23} or
confluent hypergeometric functions\cite
{GK61,B88a,WGMD02,PR03,DSL05,IS07b,OAA08,AS12,PD13,BBSM23}. The radial
eigenvalue equation can also be solved by means of the power-series method%
\cite{SG85} or the closely related wavefunction ansatz\cite{IS07a,IS08}.
Other authors proposed the application of a variety of algebraic methods\cite
{B88b,BS90,BDO92,L94,P98b,DM02,S03,DSL05,SD06,BB22} as well as supersymmetry%
\cite{KGS90,PRML05}, the quantization approach\cite{KGS90,DMG07},
path integral\cite{SE88} and the Nikiforov-Uvarov
method\cite{STAY07,TS09}. The exact solutions of the
Schr\"{o}dinger equation for the singular harmonic oscillator
proved suitable for the treatment of other problems\cite
{S84,R05,SHK05} including the application of perturbation
theory\cite{SHV03}. There has also been interest in the
thermodynamic functions for a gas of pseudoharmonic
molecules\cite{BD90,P95,B88a,P01}, in the virial and
Hellmann-Feynman theorems\cite{B88a,P98a,P99,PS07,PS07,OAA08,OS12}
as well as in the coherent states\cite{P98b,P01}.

From a pedagogical point of view, Palma and Raff\cite{PR03} suggested the
one-dimensional harmonic oscillator in the presence of a dipole-like
interaction. This model is a particular case of the singular harmonic
oscillator and has been treated by Pimentel and de Castro\cite{PD13} with
somewhat more detail. Palma and Raff\cite{PR03} and Pimentel and de Castro%
\cite{PD13} wrote the potential-energy function of the singular harmonic
oscillator as $V(x)=m\omega ^{2}x^{2}/2+\hbar ^{2}\alpha /\left(
2mx^{2}\right) $ where $\alpha $ is a dimensionless parameter related to the
strength of the singular term. The former authors considered only the case $%
\alpha >0$ and the latter also negative values of this parameter. Since the
discussion of the behaviour of the eigenfunctions at origin for this
quantum-mechanical model appears to be of great pedagogical interest, a
further analysis may be worthwhile.

In this paper we discuss the singular harmonic oscillator with somewhat more
detail. In section~\ref{sec:model} we outline the model and the behaviour of
the eigenfunctions at origin. In section~\ref{sec:Frobenius} we apply the
power-series method and obtain the eigenvalues and eigenfunctions. A
discussion about the existence of allowed values of the energy
(quantization) is given in appendix~\ref{sec:bounds}. Finally in section~\ref
{sec:conclusions} we summarize the main results and draw conclusions.

\section{The model}

\label{sec:model}

The pseudoharmonic potential has been expressed in several different ways;
for example:
\begin{equation}
V(r)=\frac{kr_{0}^{2}}{8}\left( \frac{r_{0}}{r}+\frac{r}{r_{0}}\right)
^{2}=D_{0}\left( \frac{r_{0}}{r}+\frac{r}{r_{0}}\right) ^{2}=D_{e}\left(
\frac{r_{e}}{r}+\frac{r}{r_{e}}\right) ^{2}=\frac{A}{r^{2}}+Br^{2}+C.
\label{eq:V_SHO}
\end{equation}
In passing, we mention that some authors considered $D_{0}$ ($D_{e}$) to be
the dissociation energy between two atoms in a solid\cite{IS07b,IS08,OS12}.
In the first place, it makes no sense to speak of an isolated pair of atoms
in a solid. Note that those authors applied the model to isolated diatomic
molecules (gas phase). Second, the pseudoharmonic potential, as any infinite
square well, does not allow dissociation. Perhaps, the origin of the latter
misunderstanding lays in the fact that this potential has been mistook for
the\ Kratzer one\cite{SD06} (or considered of Kratzer type\cite{BS90}) that
already exhibits dissociation energy.

In this paper we consider the time-independent Schr\"{o}dinger equation with
the Hamiltonian operator
\begin{equation}
H=-\frac{\hbar ^{2}}{2m}\frac{d^{2}}{dx^{2}}+\frac{V_{-2}}{2x^{2}}+\frac{%
V_{2}}{2}x^{2},\;x>0,  \label{eq:H}
\end{equation}
where $m$ is the mass of the particle (or the reduced mass of a pair of
particles) and $V_{2}>0$. We will discuss possible values of $V_{-2}$ later
on. It is convenient for present purposes to restrict the domain to positive
values of the coordinate $x$. If we choose the units of length $L=\hbar
^{1/2}/\left( mV_{2}\right) ^{1/4}$ and energy $\hbar \sqrt{V_{2}/m}=\hbar
\omega $ then we obtain the dimensionless Hamiltonian operator\cite{F20}
\begin{equation}
\tilde{H}=\frac{1}{\hbar \omega }H=-\frac{1}{2}\frac{d^{2}}{d\tilde{x}^{2}}+%
\frac{\alpha }{2\tilde{x}^{2}}+\frac{1}{2}\tilde{x}^{2},\;\alpha =\frac{%
mV_{-2}}{\hbar ^{2}},\;\tilde{x}=\frac{x}{L}.  \label{eq:H_dim}
\end{equation}
Note that the Hamiltonian operator in equation (\ref{eq:H}) is identical to
that of Palma and Raff\cite{PR03} and Pimentel and de Castro\cite{PD13}
because $V_{-2}=\hbar ^{2}\alpha /m$ and $V_{2}=m\omega ^{2}$ as shown by
the above equations. However, it is more practical to carry out the
calculation with the dimensionless expression (\ref{eq:H_dim}). From now on,
we omit the tilde over the dimensionless observables $\tilde{H}$ and $\tilde{%
x}$.

We first consider the behaviour of the eigenfunctions at origin that is
dominated by the singular term $\alpha /(2x^{2})$; therefore, we focus on
the operator
\begin{equation}
H_{s}=-\frac{1}{2}\frac{d^{2}}{dx^{2}}+\frac{\alpha }{2x^{2}},
\label{eq:H_s}
\end{equation}
because $x^{-2}\gg x^{2}$ when $x\ll 1$. If we choose $\varphi (x)=x^{s}$
then we have $H_{s}\varphi =0$ provided that $\alpha =s(s-1)$. The two roots
of this equation are
\begin{equation}
s_{\pm }=\frac{1}{2}\left( 1\pm \sqrt{1+4\alpha }\right) .  \label{eq:s_+-}
\end{equation}
$\varphi (x)$ is regular at origin if $s>0$; therefore, $\alpha \geq -1/4$.
If $-1/4<\alpha \leq 0$ then $1\geq s_{+}>1/2>s_{-}\geq 0$ and both roots
are suitable if we only require that $\varphi \left( 0^{+}\right) $ be
finite. If $\alpha >0$ then $s_{+}>0>s_{-}$ and only $s_{+}$ is acceptable.
When $\alpha =-1/4$ the two exponential solutions are identical because $%
s_{\pm }=1/2$. In order to obtain the second solution we proceed as follows:
\begin{eqnarray}
\frac{\partial }{\partial s}\left( H_{s}\varphi \right) &=&H_{s}\frac{%
\partial }{\partial s}\varphi +\frac{2s-1}{2}x^{s-2}=0,  \nonumber \\
\left. \frac{\partial }{\partial s}\left( H_{s}\varphi \right) \right|
_{s=1/2} &=&H_{s}\left. \frac{\partial }{\partial s}\varphi \right|
_{s=1/2}=0,
\end{eqnarray}
where we substituted $s(s-1)$ for $\alpha $ in $H_{s}$. In this way, we
realize that the two solutions for $\alpha =-1/4$ are $\varphi _{1}=x^{1/2}$
and $\varphi _{2}=x^{1/2}\ln x$\cite{PD13}. It is clear that $\varphi
^{\prime }(x)$ diverges as $x\rightarrow 0^{+}$ when $-1/4\leq \alpha <0$
but it is no reason for rejecting this kind of solutions.

\section{Frobenius method}

\label{sec:Frobenius}

Many textbooks on quantum mechanics and quantum chemistry solve the
Schr\"{o}dinger equation for the harmonic oscillator and hydrogen atom
(among other exactly-solvable problems) by means of the power-series method%
\cite{CDL77,P68}. It is probably the simplest and most intuitive approach
for introductory courses on quantum mechanics and quantum chemistry and in
what follows we apply it to the dimensionless singular harmonic oscillator
\begin{equation}
H=T+V,\;T=-\frac{1}{2}\frac{d^{2}}{dx^{2}},\;V(x)=\frac{\alpha }{2x^{2}}+%
\frac{1}{2}x^{2}.  \label{eq:H_dim_2}
\end{equation}

If we assume that $\alpha >-1/4$ we can try solutions of the form
\begin{equation}
\psi (x)=x^{s}e^{-x^{2}/2}u(x),\;u(x)=\sum_{j=0}^{\infty }c_{j}x^{2j},
\label{eq:psi_series}
\end{equation}
where the factor $x^{s}$ takes into account the behaviour of $\psi
(x)$ at origin already discussed in section~\ref{sec:model} and
$e^{-x^{2}/2}$ comes from the behaviour of the wavefunction at
infinity, determined by the harmonic term. We are looking for
bound states given by square-integrable solutions:
\begin{equation}
\int_{0}^{\infty }\left| \psi (x)\right| ^{2}dx<\infty .
\label{eq:normalization}
\end{equation}
Upon substituting the ansatz (\ref{eq:psi_series}) into the dimensionless
Schr\"{o}dinger equation $H\psi =E\psi $ we can easily verify that the
coefficients $c_{j}$ satisfy the two-term recurrence relation
\begin{eqnarray}
c_{j+1} &=&A_{j}c_{j},\;j=0,1,\ldots ,\;c_{0}\neq 0,  \nonumber \\
A_{j} &=&\frac{4j+2s+1-2E}{2\left( j+1\right) \left( 2j+2s+1\right) }.
\label{eq:TTRR}
\end{eqnarray}
It is not difficult to prove that $A_{j}>\beta /(j+1)$, $1/2<\beta <1$, when
$j>\frac{E}{2\left( 1-\beta \right) }+\frac{\left( 2s+1\right) \left( 2\beta
-1\right) }{4\left( 1-\beta \right) }$. Therefore, it follows from the
results of the appendix~\ref{sec:bounds} that $u(x)>C\exp \left( \beta
x^{2}\right) +P_{k}(x^{2})$, where $C>0$ and $P_{k}(x^{2})$ is a polynomial
function of $x^{2}$ of degree $k$. We thus conclude that $\psi (x)$ is not
square integrable unless
\begin{equation}
E=E_{n,s}=2n+s+\frac{1}{2},\;n=0,1,\ldots ,  \label{eq:E_(n.s)}
\end{equation}
from which it follows that
\begin{equation}
A_{j}=\frac{2\left( j-n\right) }{\left( j+1\right) \left( 2j+2s+1\right) }.
\label{eq:A_j_n}
\end{equation}
Note that $c_{j}=0$ for all $j>n$ when $E$ is one of the allowed values $%
E_{n,s}$ of the energy and the series $u(x)$ reduces to a polynomial of
degree $2n$. In other words, the eigenfunctions are square integrable
because they are of the form
\begin{equation}
\psi _{n,s}(x)=x^{s}e^{-x^{2}/2}\sum_{j=0}^{n}c_{j,s}x^{2j}.
\label{eq:psi_(n,s)}
\end{equation}
We think that this simple analysis is suitable for pedagogical purposes. It
is supposed to be somewhat more rigorous than the arguments that appear in
most textbooks\cite{P68} and used by Sage and Goodisman\cite{SG85} for the
pseudoharmonic oscillator.

\begin{figure}[tbp]
\begin{center}
\includegraphics[width=9cm]{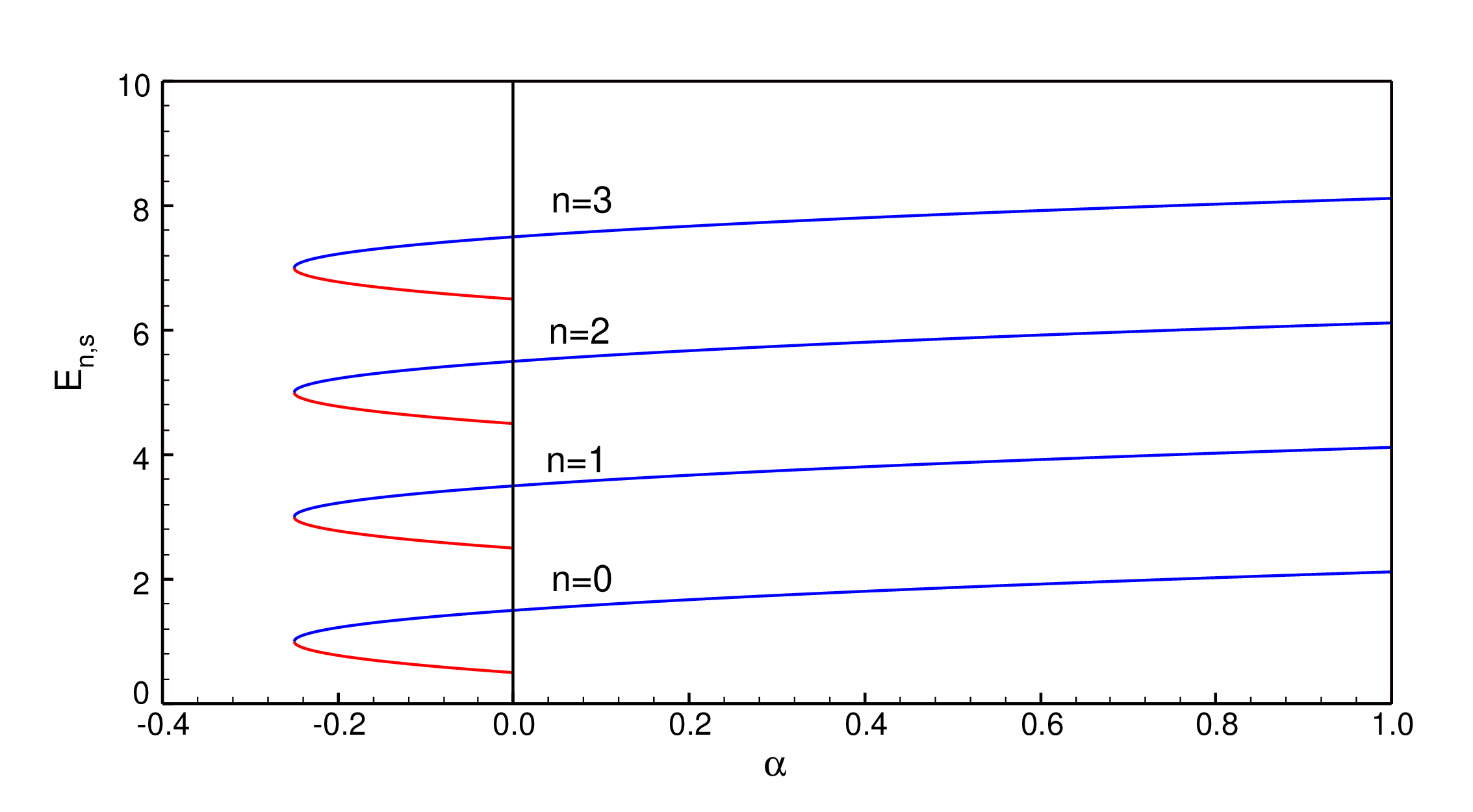}
\end{center}
\caption{First four eigenvalues of the dimensionless singular harmonic
oscillator }
\label{Fig:Ens}
\end{figure}

Figure~\ref{Fig:Ens} shows the first four eigenvalues $E_{n,s}$ for some
values of $\alpha $. We appreciate that the number of eigenvalues and
eigenfunctions for $-1/4<\alpha \leq 0$ is twice the number of those for $%
\alpha >0$. The reason is that the solution with $s=s_{-}<0$ for $\alpha >0$
is not regular at origin and, consequently, it is not acceptable from a
physical point of view. Palma and Raff\cite{PR03} did not consider the case $%
\alpha <0$ and Pimentel and de Castro\cite{PD13} only showed those solutions
with $s=s_{+}$ in their figures. For this reason, we think that our figure~%
\ref{Fig:Ens} may be revealing from a pedagogical point of view. For $\alpha
=0$ the solutions with $s_{-}=0$ and $s_{+}=1$ give rise to the well known
even and odd states, respectively, of the harmonic oscillator if we consider
the variable interval $-\infty <x<\infty $\cite{CDL77,P68}. Present figure~%
\ref{Fig:Ens} shows that the eigenvalues for these even and odd states come
from the two branches $E_{n,s_{+}}$ and $E_{n,s_{-}}$ as $\alpha \rightarrow
0^{-}$. Pimentel and de Castro explicitly indicated the eigenvalues for the
even and odd states in their figure~4 but did not show where the former come
from.

The behaviour of the eigenvalues for $-1/4<\alpha \leq 0$ is most
interesting. According to the Hellmann-Feynman theorem (HFT)\cite{G32,F39}
(see\cite{CDL77,P68} for a more pedagogical introduction) we have
\begin{equation}
\frac{\partial E}{\partial \alpha }=\left\langle \frac{1}{x^{2}}%
\right\rangle >0,  \label{eq:HFT}
\end{equation}
which is obviously satisfied by the solutions with $s=s_{+}$ (blue lines)
but not by those with $s=s_{-}$ (red lines). The reason is that this theorem
does not apply to the latter solutions because $0<s_{-}<1/2$ and,
consequently, the expectation value in equation (\ref{eq:HFT}) is divergent.
Ballhausen\cite{B88a} first considered the solutions with $s_{-}$ to be
physically acceptable, but after Senn's judicious criticism\cite{S89} he
changed his mind\cite{B89} and added that such solutions did not give a
finite value of $\left\langle T\right\rangle $. As argued above, such
solutions do not give a finite expectation value for $x^{-2}$.

The solutions with $s_{+}$ satisfy the virial theorem $2\left\langle
T\right\rangle =\left\langle xV^{\prime }\right\rangle $ which, together
with $E=\left\langle T\right\rangle +\left\langle V\right\rangle $, leads to
$E=\left\langle x^{2}\right\rangle $\cite{B88a}. Although the solutions with
$s_{-}$ do not render finite values of $\left\langle T\right\rangle $, $%
\left\langle xV^{\prime }\right\rangle $ and $\left\langle V\right\rangle $
they curiously satisfy $E=\left\langle x^{2}\right\rangle $. The reason is
that the singular term cancels out in the expression $V+\frac{1}{2}%
xV^{\prime }=x^{2}$.

\begin{figure}[tbp]
\begin{center}
\includegraphics[width=9cm]{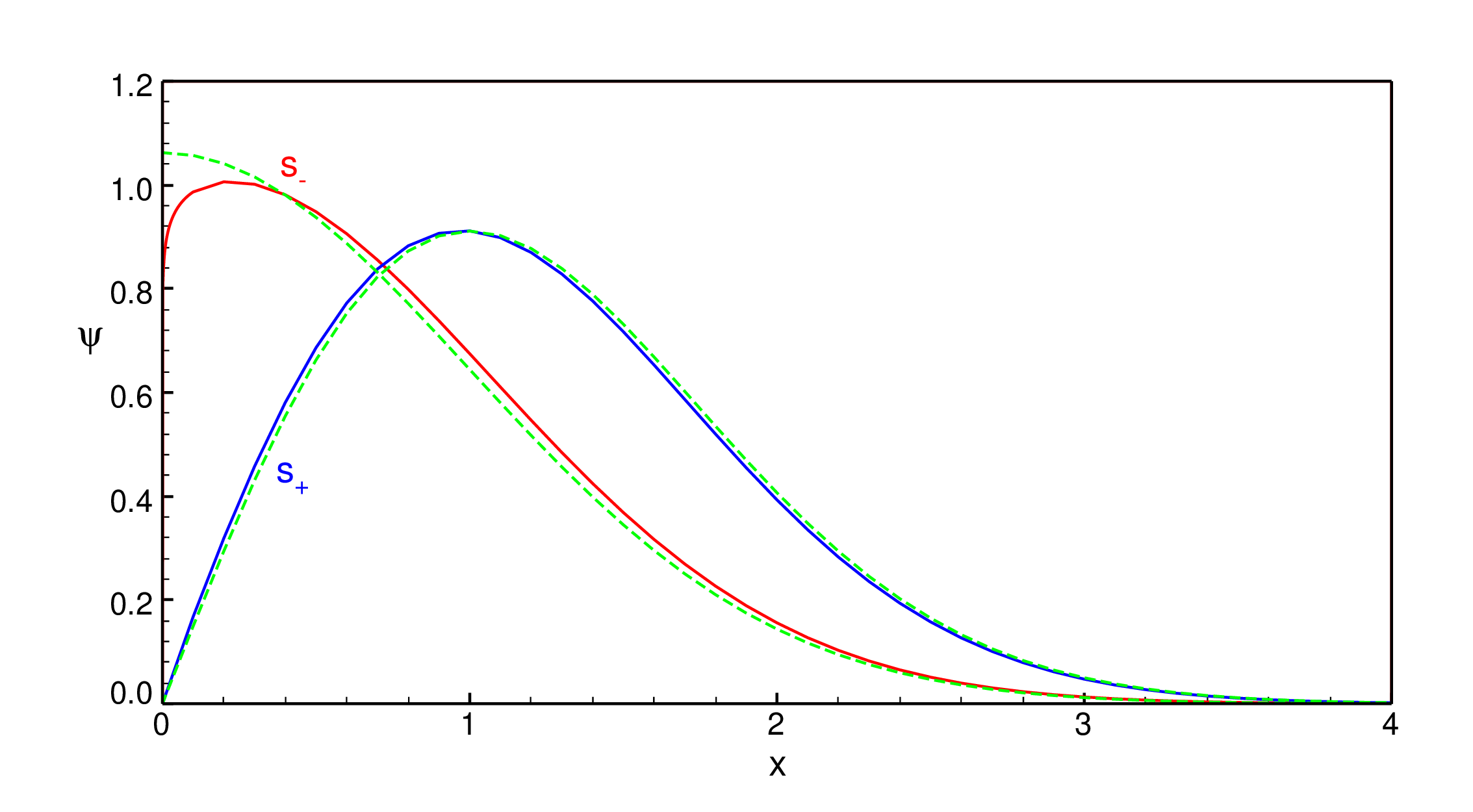}
\end{center}
\caption{Eigenfunctions $\psi_{0,s}$ for $\alpha=-0.0475$. The green, dashed
lines correspond to the eigenfunctions for $\alpha=0$}
\label{Fig:psi_0}
\end{figure}

Figure~\ref{Fig:psi_0} shows the normalized eigenfunctions
\begin{equation}
\psi _{0,s}=\sqrt{\frac{2}{\Gamma \left( s+1/2\right) }}x^{s}e^{-x^{2}/2},
\label{eq:psi_(0,s)}
\end{equation}
for $\alpha =-0.0475$. Both solutions with $s=s_{-}=0.05$ and $s=s_{+}=0.95$
are square integrable; however, the former does not satisfy the HFT as
indicated above. This figure also shows the harmonic-oscillator
eigenfunctions $\psi _{0,0}$ and $\psi _{0,1}$ ($\alpha =0$, green dashed
lines) for comparison. Since $\left| \alpha \right| \ll 1$ this figure shows
that the eigenfunctions of the singular harmonic oscillator approach those
of the well known harmonic oscillator as $\alpha \rightarrow 0^{-}$. It is
worth pointing out that all the solutions just mentioned are normalized in
the interval $0<x<\infty $.

\section{Conclusions}

\label{sec:conclusions}

In this paper we enlarged the results of Parma and Raff\cite{PR03} and
Pimentel and de Castro\cite{PD13} by considering both solutions of the
Schr\"{o}dinger equation for the attractive singular potential. In this way
we derived two branches of the eigenvalues in the range $-1/4<\alpha <0$
that were not considered in those papers. An interesting outcome of this
analysis is that one of the branches ($s=s_{+}$) satisfies the HFT (\ref
{eq:HFT}) while the other ($s=s_{-}$) does not. That the solution to a
Schr\"{o}dinger equation may not satisfy such celebrated theorem may be
revealing in a course on quantum mechanics. This fact is most important for
a discussion of the conditions under which a mathematical result is derived.
Another interesting feature of present discussion is that it shows that the
even and odd eigenfunctions of the harmonic oscillator stem from the
solutions of the singular harmonic oscillator with $s=s_{+}$ and $s=s_{-}$,
respectively.

Both Parma and Raff\cite{PR03} and Pimentel and de Castro\cite{PD13}
obtained the eigenvalues from the properties of the confluent hypergeometric
function. In order to follow such an analysis the students should resort to
suitable books or tables of functions and integrals in order to make use of
the asymptotic behaviour of such solutions. In the present case, on the
other hand, we resorted to the Frobenius (power-series) method that the
students may probably find more intuitive, clearer and easier to follow from
beginning to end. This approach can obviously be applied to any
quantum-mechanical problem in which the solution of the Schr\"{o}dinger
equation can be reduced to a two-term recurrence relation for the
coefficients of the series expansion. It is clear that a teacher should also
encourage the students to resort to well known functions and polynomials
that may be found in available literature on applied mathematics, but we
think that the comparison of both strategies may be most fruitful.

In addition to the simple argument for truncating the power-series given in
most textbooks\cite{P68} (see also\cite{SG85}), appendix\ref{sec:bounds}
provides a somewhat more rigorous argument that in our opinion is suitable
for a course on quantum mechanics or quantum chemistry.

\appendix

\numberwithin{equation}{section}

\section{On the behaviour of a class of power series}

\label{sec:bounds}

Suppose that we have two series of the form
\begin{equation}
S(r)=\sum_{j=0}^{\infty }a_{j}r^{j},\;T(r)=\sum_{j=0}^{\infty }b_{j}r^{j},
\label{eq:series_a_b}
\end{equation}
and that their coefficients satisfy the two-term recurrence relations
\begin{eqnarray}
a_{j+1} &=&A_{j}a_{j},\;j=0,1,\ldots ,\;a_{0}\neq 0,\;  \nonumber \\
b_{j+1} &=&B_{j}b_{j},\;j=0,1,\ldots ,\;b_{0}\neq 0.  \label{eq:rec_rel_a_b}
\end{eqnarray}
We also assume that
\begin{equation}
\lim\limits_{j\rightarrow \infty }A_{j}=0,\;\lim\limits_{j\rightarrow \infty
}B_{j}=0,
\end{equation}
so that both series converge for all $r$.

If
\begin{equation}
A_{j}\geq B_{j}>0,\;j>k,
\end{equation}
then
\begin{equation}
\frac{a_{j}}{a_{k}}=A_{j-1}A_{j-2}\ldots A_{k}\geq \frac{b_{j}}{b_{k}}%
=B_{j-1}B_{j-2}\ldots B_{k},\;j>k.
\end{equation}
Without loss of generality we assume that $a_{k}>0$ and $b_{k}>0$ (note
that, if necessary, we may multiply any one of the series by $-1$ in order
to satisfy these inequalities). Therefore, for $r>0$ we have
\begin{equation}
S(r)\geq \frac{a_{k}}{b_{k}}T(r)+\sum_{j=0}^{k}\left( a_{j}-\frac{a_{k}}{%
b_{k}}b_{j}\right) r^{j}.  \label{eq:series_inequality}
\end{equation}

If $0<\lim\limits_{j\rightarrow \infty }jA_{j}=D<\infty $, then there exists
$0<\beta <D$ such that $A_{j}>\beta /(j+1)$, $j>k$, for some integer $k$.
Therefore, if we choose $B_{j}=\beta /(j+1)$ then $T(r)=e^{\beta r}$ and
\begin{equation}
S(r)\geq Ce^{\beta r}+P_{k}(r),
\end{equation}
where $C>0$ and $P_{k}(r)$ is a polynomial function of $r$ of degree $k$.
This result will prove useful for the analysis of the solutions of the
singular harmonic oscillator in section~\ref{sec:Frobenius}.


\begin{thebibliography}{99}
\bibitem{CDL77}  Cohen-Tannoudji C, Diu B, and Lalo\"{e} F 1977 \textit{%
Quantum Mechanics} (John Wiley \& Sons, New York).

\bibitem{P68}  Pilar F L 1968 \textit{Elementary Quantum Chemistry}
(McGraw-Hill, New York).

\bibitem{BO78}  Bender C M and Orszag S A 1978 \textit{Advanced mathematical
methods for scientists and engineers} (McGraw-Hill, New York).

\bibitem{D32}  Davidson P M 1932 \textit{Proc. Roy. Soc. A} \textbf{135} 459.

\bibitem{GK61}  Gol'dman I I and Krivchenkov V D 1961 \textit{Problems in
quantum mechanics} (Pergamon Press, London).

\bibitem{WJ79}  Weissman Y and Jortner J 1979 \textit{Phys. Lett. A} \textbf{%
70} 177.

\bibitem{S84}  Sage M 1984 \textit{Chem. Phys.} \textbf{87} 431.

\bibitem{SG85}  Sage M and Goodisman J 1985 \textit{Am. J. Phys.} \textbf{53}
350.

\bibitem{B88a}  Ballhausen C J 1988 \textit{Chem. Phys. Lett.} \textbf{146}
449.

\bibitem{B88b}  Ballhausen C J 1988 \textit{Chem. Phys. Lett.} \textbf{151}
428.

\bibitem{SE88}  Sever R and Erko\c{c} \c{S} 1988 \textit{Phys. Rev. A} \textbf{37}
2687.

\bibitem{S89}  Senn P 1989 \textit{Chem. Phys. Lett.} \textbf{154} 172.

\bibitem{B89}  Ballhausen C J 1989 \textit{Chem. Phys. Lett.} \textbf{154}
174.

\bibitem{BS90}  Brajamani S and Singh C A 1990 \textit{J. Phys. A} \textbf{23%
} 3421.

\bibitem{BD90}  B\"{u}y\"{u}kk\i l\i\c{c} F and Demirhan D 1990 \textit{Chem. Phys.
Lett.} \textbf{166} 272.

\bibitem{KGS90}  Kasap E, G\"{o}n\"{u}l B, and \c{S}im\c{s}ek M 1990 \textit{Chem.
Phys. Lett.} \textbf{172} 499.

\bibitem{BDO92}  B\"{u}y\"{u}kk\i l\i\c{c} F, Demirhan D, and \"{O}zeren F 1992
\textit{Chem. Phys. Lett.} \textbf{194} 9.

\bibitem{L94}  L\'{e}vai G 1994 \textit{J. Phys. A} \textbf{27} 3809.

\bibitem{P95}  Popov D 1995 \textit{Acta Phys. Slov.} \textbf{45} 557.

\bibitem{P98a}  Popov D 1998 \textit{Int. J. Quantum Chem.} \textbf{69} 159.

\bibitem{P98b}  Popov D 1998 \textit{Acta Phys. Slov.} \textbf{48} 1.

\bibitem{P99}  Popov D 1999 \textit{Czech. J. Phys.} \textbf{49} 1121.

\bibitem{P01}  Popov D 2001 \textit{J. Phys. A} \textbf{34} 5283.

\bibitem{WGMD02}  Wang L-W, Gu X-Y, Ma Z-Q, and Dong S-H 2002 \textit{Found.
Phys. Lett.} \textbf{15} 569.

\bibitem{DM02}  Dong S-H and Ma Z-Q 2002 \textit{Int. J. Mod. Phys. E}
\textbf{11} 155.

\bibitem{PR03}  Palma G and Raff U 2003 \textit{Am. J. Phys.} \textbf{71}
247. Addendum: Am. J. Phys. \textbf{71} , 956 (2003)

\bibitem{S03}  Dong S-H 2003 \textit{Appl. Math. Lett.} \textbf{16} 199.

\bibitem{SHV03}  Saad N, Hall R L, and von Keviczy A B 2003 \textit{J. Phys.
A} \textbf{36} 487.

\bibitem{DSL05}  Dong S-H, Sun G-H, and Lozada-Cassou M 2005 \textit{Int. J.
Mod. Phys. A} \textbf{20} 5663.

\bibitem{PRML05}  Pe\~{n}a J J, Romero-Romo M A, Morales J, and
L\'{o}pez-Bonilla J L 2005 \textit{Int. J. Quantum Chem.} \textbf{105} 731.

\bibitem{R05}  Rowe D J 2005 \textit{J. Phys. A} \textbf{38} 10181.

\bibitem{SHK05}  Saad N, Hall R L, and Katatbeh Q, D. 2005 \textit{J. Math.
Phys.} \textbf{46} 022104.

\bibitem{SD06}  Singh C A and Devi O B 2006 \textit{Int. J. Quantum Chem.}
\textbf{106} 415.

\bibitem{STAY07}  Sever R, Tezcan C, Akta\c{s} M, and Ye\c{s}ita\c{s} \"{O} 2007 \textit{%
J. Math. Chem} \textbf{43} 845.

\bibitem{PS07}  Patil S H and Senn K D 2007 \textit{Phys. Lett. A} \textbf{%
362} 109.

\bibitem{DMG07}  Dong S-H, Morales D, and Garc\'{i}a-Ravelo J 2007 \textit{%
Int. J. Mod. Phys. E} \textbf{16} 189.

\bibitem{IS07a}  Ikhdair S M and Sever R 2007 \textit{Cent. Eur. J. Phys.}
\textbf{5} 516.

\bibitem{IS07b}  Ikhdair S M and Sever R 2007 \textit{J. Molec. Struct.
(THEOCHEM)} \textbf{806} 155.

\bibitem{IS08}  Ikhdair S M and Sever R 2008 \textit{Cent. Eur. J. Phys.}
\textbf{6} 697.

\bibitem{OAA08}  Oyewumi K J, Akinpelu F O, and Agboola A D 2008 \textit{%
Int. J. Theor. Phys.} \textbf{47} 1039.

\bibitem{C08}  \c{C}etin A 2008 \textit{Phys. Lett. A} \textbf{372} 3852.

\bibitem{TS09}  Tezcan C and Sever R 2009 \textit{Int. J. Quantum Chem.}
\textbf{48} 337.

\bibitem{AS12}  Arda A and Sever R 2012 \textit{J. Math. Chem} \textbf{50}
971.

\bibitem{OS12}  Oyewumi K J and Senn K D 2012 \textit{J. Math. Chem} \textbf{%
50} 1039.

\bibitem{PD13}  Pimentel D R M and de Castro A S 2013 \textit{Revista
Brasileira de Ensino de F\'{i}sica} \textbf{35} 3303.

\bibitem{ND16}  Nogeira P H F and de Castro A S 2016 \textit{J. Math. Chem}
\textbf{54} 1783.

\bibitem{BB22}  Baykal M and Baykal A 2022 \textit{Eur. J. Phys.} \textbf{43}
035406.

\bibitem{BBSM23}  Boyack R, Bhuiyan A, Su A, and Marsiglio F 2023 \textit{J.
Math. Chem} \textbf{61} 242.

\bibitem{F20}  Fern\'{a}ndez F M 2020 Dimensionless equations in
non-relativistic quantum mechanics. arXiv:2005.05377 [quant-ph]

\bibitem{G32}  G\"{u}ttinger P 1932 \textit{Z. Phys.} \textbf{73} 169.

\bibitem{F39}  Feynman R P 1939 \textit{Phys. Rev.} \textbf{56} 340.
\end{thebibliography}
\end{document}